\documentclass[10pt,conference]{IEEEtran}
\IEEEoverridecommandlockouts

\usepackage{cite}
\usepackage{amsmath,amssymb,amsfonts}
\usepackage{algorithmic}
\usepackage{graphicx}
\usepackage{textcomp}
\usepackage{xcolor}
\usepackage{url}
\usepackage{comment}
\usepackage[caption=false,font=footnotesize]{subfig}
\def\BibTeX{{\rm B\kern-.05em{\sc i\kern-.025em b}\kern-.08em
    T\kern-.1667em\lower.7ex\hbox{E}\kern-.125emX}}
\begin{document}

\title{Tracybot: Tracing LLM-Generated Code to Prompts}

\title{Tracybot: Preserving Developer Intent by Tracing Agent-Generated Code to Prompts}

\author{\IEEEauthorblockN{Răzvan Albu}
\IEEEauthorblockA{
\textit{University of Gothenburg}\\
Gothenburg, Sweden \\
0009-0000-9470-6204}
\and
\IEEEauthorblockN{Aliaksei Khval}
\IEEEauthorblockA{
\textit{University of Gothenburg}\\
Gothenburg, Sweden \\
0009-0001-2356-0815}
\and
\IEEEauthorblockN{Maxine Orlova}
\IEEEauthorblockA{
\textit{University of Gothenburg}\\
Gothenburg, Sweden \\
0009-0002-2227-0557}
\and
\IEEEauthorblockN{Danis Music}
\IEEEauthorblockA{
\textit{University of Gothenburg}\\
Gothenburg, Sweden \\
0009-0006-2443-8009}
\and
\IEEEauthorblockN{Filipe Araújo Veras Rosa}
\IEEEauthorblockA{
\textit{University of Gothenburg}\\
Gothenburg, Sweden \\
0009-0000-3371-3443}

\and
\IEEEauthorblockN{Lirong Yi}
\IEEEauthorblockA{
\textit{Chalmers University of Technology}\\
\textit{and University of Gothenburg}\\
Gothenburg, Sweden \\
0009-0008-7536-8367}

\and
\IEEEauthorblockN{Ranim Khojah}
\IEEEauthorblockA{
\textit{Chalmers University of Technology}\\
\textit{and University of Gothenburg}\\
Gothenburg, Sweden \\
0000-0002-1090-3153}
}

\maketitle

\begin{abstract}
Large language models (LLMs) are now widely used in software development, and LLM-based coding agents are given increasing autonomy. With coding agents, many design decisions are made in the conversation between the developer and the agent. 
Typically, the developer starts a conversation with an agent and states their intent, the agent proposes options if needed, and the developer plans the change accordingly. Then, the resulting code is committed.
However, in such scenarios, the decisions and the intent behind the changes remain in the conversation, with no link to the code they produced. 
In this paper, we present Tracybot, an open-source tool that integrates with VS Code, Git, and coding agents to keep track of relevant conversations in an LLM-assisted development process. 
Tracybot captures each change as a \textit{Tasklet} which consists of optional planning prompts used to plan a change, together with a build prompt that produces it. 
Tracybot has two main features: (i) it maps line-level code changes made by coding agents to the Tasklets that produced them, and (ii) it allows the developer to export a Tasklet as a decision record when needed, for example when the decision concerns the architecture of the project. We developed the tool over two iterations of a design science study, while focusing the evaluation on the usability and user-friendliness of the tool. 
In a focus group (7 participants) and a survey (24 participants), we found that Tracybot was perceived as useful for activities like code review, and most participants found our Tasklet design and presentation to be a clear way to represent the developer's intent. 
Tool: \url{https://github.com/TracyTeam/tracybot}. Video: \url{https://youtu.be/Vg_a1F5eO04}.

\end{abstract}
 
\begin{IEEEkeywords}
prompt traceability, developer intent, decision records, coding agents
\end{IEEEkeywords}
 
\section{Introduction}
 
The use of LLM coding agents is becoming a standard part of modern software development. However, the adoption of such agents has outpaced the practices around them, especially in software companies~\cite{khojah2026}.
Few developers explicitly credit AI tools for the code they generate~\cite{kraishan2025}, especially after editing the output~\cite{kashif2025}.
Therefore, regulations such as the EU AI Act~\cite{euaiact} push companies towards clear AI crediting, and software companies believe prompts used during development need to be stored somehow~\cite{khojah2026}.

Crediting AI usually means labelling code as AI-generated, but a label does not say \textit{why} the code was written; the prompt does.
In an eye-tracking study, practitioners changed their reviewing behaviour when reviewing code labelled as AI-generated, and some used the prompt to guide their review, which calls the need to treat the prompt as a software artifact~\cite{khojah2026eye}.

The prompt matters even more because of how developers now work with agents. Many implementation and design decisions are made in conversation with the agent.
Developers typically start by explaining what they want along with some constraints. The agent then suggests a plan, and they refine it together over several prompts~\cite{tang2026specstory}. During such conversations, decisions are made, and eventually the developer requests the agent to apply the planned changes in the codebase. 
While the change is made and the code is then committed, the reasoning behind it is not. It stays in the developer's conversation history, where teammates cannot see it and where it is easily lost.

Storey calls the absence of explicit goals, constraints, and rationale \textit{intent debt}, and notes that intent is best documented at the time a decision is made, since recovering it later can be difficult~\cite{storey2026}. 
One way of minimising the risk of intent debt is the use of decision records, which state what was decided, why, and what was ruled out. Conversations with agents contain exactly this information. 

To close these gaps, we present Tracybot, an open-source tool that integrates with the developer's IDE (VS Code), coding agents (Claude, Codex, and OpenCode), and Git. Tracybot records each change discussed in a conversation with an agent as a \textit{Tasklet}, which groups several prompts planning the change with a final build prompt that applies the change in the codebase, together with the agent's responses. 

Its main features are:
(i)~automatically capturing Tasklets during development and \textbf{mapping prompts to the LLM-generated code}, so that developers can click a line in the editor to read the conversation that led to it;
(ii)~\textbf{keeping a Tasklet history} of earlier Tasklets that changed the same line, so that later changes do not hide the reasoning behind earlier ones, and so that developers can see how the generated code evolved;
(iii)~ \textbf{maintaining the attribution} as the code changes, so that small edits by the developer (e.g., formatting) leave the AI attribution in place, while a major human rewrite (e.g., changing the entire logic) removes it; and
(iv)~\textbf{exporting Tasklets as decision records} on demand to turn important decisions in conversations into standalone documents.
All of this runs in the background using Git features, without interfering with how developers prompt or commit.

\section{Background and Existing Tools}
 
In software engineering, traceability is the ability to link related artifacts, such as a requirement and the code that implements it, so that one can be found from the other~\cite{gotel2012}.
In LLM-assisted development, prompts play a similar role to requirements, as they can drive software development~\cite{khojah2026eye}.

Recent tools have started to bring traceability to AI-generated code (Table~\ref{tab:tools}). 
Git~AI\footnote{\url{https://github.com/git-ai-project/git-ai}} and Cursor Blame\footnote{\url{https://cursor.com/docs/integrations/cursor-blame}} (Enterprise plan) extend \texttt{git blame} with AI attribution. They record lines of code an agent wrote, link them to the session~\footnote{An agent session contains a conversation with all interactions from when the agent is started until it is closed.} that produced them, and show a short summary when the developer hovers over a line. PromptTrace\footnote{VS Code Marketplace, \texttt{ganeshutla.prompt-trace}, v0.0.6.} treats each prompt separately, where it links every prompt to the diff it produced and each line to the single prompt that introduced it.

\begin{table}[!ht]
\caption{Tools that link code to prompts (as of October 2026).}
\label{tab:tools}
\centering
\footnotesize
\begin{tabular}{@{}p{1.57cm}p{1.33cm}p{1.92cm}p{2.8cm}@{}}
\hline
\textbf{Tool} & \textbf{In editor} & \textbf{Unit linked} & \textbf{What the user sees} \\
\hline
Git AI & Yes & Agent session & Prompts summary \\
Cursor Blame & Cursor only & Agent session & Session summary \\
PromptTrace & Yes & Single prompt & Prompt and its diff \\
\textbf{Tracybot} & Yes & Tasklet (plan and build prompts) & Prompts and responses, lines it touched, and earlier Tasklets for the line \\
\hline
\end{tabular}
\end{table}
 
These tools are designed for AI attribution: knowing how much of the code an agent wrote. Tracybot is designed to preserve the decisions behind that code (made by the developer of the agent), and differs from them in four ways.

First, it links a line to a Tasklet. Git~AI and Cursor Blame link a line to a whole session, which may cover many unrelated tasks, while PromptTrace links it to a single prompt, which can be uninformative when it is a short instruction such as ``go ahead'' that relies on a plan in a previous prompt. A Tasklet contains exactly the planning and the build prompt behind one change.
Second, it shows the full prompts and responses, not a summary. A summary leaves out the options that were considered and the reasons for the choice.
Third, it keeps the history of Tasklets for each line, so later changes to the same code do not hide earlier reasoning.
Fourth, it lets developers export a Tasklet as a decision record, which we did not find in these tools.
To our knowledge, no other tool combines these features in the editor. 

\begin{figure*}[!ht]
\centering
    \includegraphics[width=0.95\linewidth]{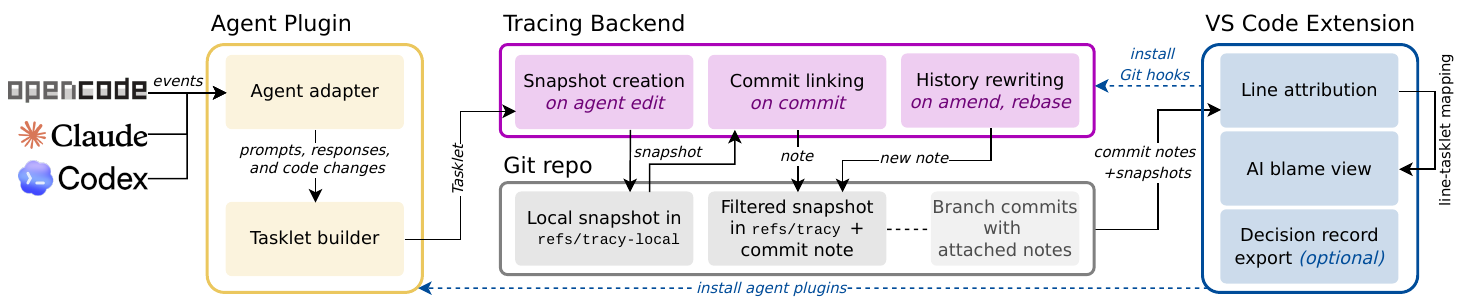}
\caption{Architecture of Tracybot.}
\label{fig:arch}
\end{figure*}

\section{Tracybot}

Tracybot has three components (Fig.~\ref{fig:arch}): agent plugins, a tracing backend, and a VS Code extension. The tracing logic is separate from the plugins and the extension, so supporting another agent or editor only requires a thin adapter.

\subsection{Agent Plugins}
\label{sec:plugins}

Tracybot supports OpenCode, Claude Code, and Codex, with one plugin per agent. Each plugin has two parts: an \textbf{agent adapter}, which connects to the agent and records what happens during a session, and a \textbf{Tasklet builder}, which groups the recorded conversation and passes it, together with the code changes, to the tracing backend. The plugins run silently with no action required from the developer.

The \textbf{agent adapter} depends on what each agent offers for extensions. OpenCode loads plugins into its own process and exposes the full session through its SDK, so the adapter can read every message directly. Claude Code and Codex instead run \emph{hooks}, i.e., short external commands triggered at fixed events, such as after a file is edited or when the agent finishes a reply. Their adapters therefore read the prompts from the agent's transcript and the response from the final event. To extend Tracybot support to other agents, relevant Plugin APIs or hooks can be added to this component.

The \textbf{Tasklet builder} decides which parts of the conversation belong together. Developers often discuss a change with the agent over several prompts before asking it to apply it, so the final request is often a short instruction like \textit{``Go with the first option''}. Recording only that request would lose the prompts that show the developer's intent. The builder therefore groups the conversation into \emph{Tasklets}. A Tasklet consists of the prompt that changed the code, i.e., the \emph{build} prompt, together with the \emph{planning} prompts that preceded it without changing any code, the agent's responses, and the resulting code changes. For OpenCode, the builder uses the agent's Plan and Build modes to separate the two. Claude Code and Codex have no such modes, so their builders treat up to five earlier prompts in the same session that changed no files as planning prompts.

\subsection{Tracing Backend}

The backend consists of a Python script, which the agent plugins call after each change, and a set of Git hooks. Together they handle three events: an agent edit, a commit, and a rewrite of the history (when needed).

\textbf{Snapshot creation.} Right after the agent changes files, the backend stores the state of the tracked files as a hidden snapshot commit under \texttt{refs/tracy-local/*}. A snapshot of the developer's state is taken just before the agent's edits and at each pre-commit hook, so an agent snapshot contains only the agent's changes. The snapshots form a chain that never appears in the branch history, so the developer's commits, \texttt{git log}, and \texttt{git blame} stay exactly as they would be without Tracybot. The chain is also kept local, because it may contain changes that are never committed, such as experiments the developer later discards or files that should not be shared; only the snapshots of committed files are shared later.

\textbf{Commit linking.} When the developer commits, the post-commit hook links the new commit to its snapshot chain with a Git note\footnote{A Git note is a short text attached to a commit. Notes are stored separately from the commit, so adding one does not change the commit or its hash.}. Before sharing the chain, the hook filters it by keeping only the snapshots of files that are part of the commit, dropping snapshots that touched none of them, and storing the result under \texttt{refs/tracy}. The raw local chain is then removed. Only \texttt{refs/tracy} and the notes are pushed to the remote, so files that the developer never committed (e.g., including local experiments or sensitive data) are not shared.

\textbf{History rewriting.} When a commit is amended or rebased, a post-rewrite hook merges the tracking chains and attaches a new note to the rewritten commit, so it stays linked to its original snapshots. Further hooks synchronise the notes with the remote, so those who pull the branch see the same trace.

\subsection{VS Code Extension}
\label{sec:ext}

The extension is the part of Tracybot that developers interact with through Visual Studio Code (VS Code). The extension checks if the project opened is a Git project, then it installs the Git hooks of the tracing backend and the plugins of the detected agents. 
The extension has three main functionalities:

\textbf{Line attribution.}
To find out which Tasklet wrote each line, the extension first rebuilds the history of the file from the normal commits, the Git notes, and the snapshots from the Git repository. 
It then follows each line through this history, filters out the changes made by the developer, and maps the remaining LLM-generated lines to their Tasklets. 
Since the local snapshots are also accessible through the refs, the line-to-Tasklet mapping is available even before the developer commits.
In addition, small edits made by the developer, such as formatting or renaming, should not change the attribution from agent to human. Therefore, the extension compares the old and new version of each changed block (Git hunk) with the BLEU similarity score~\cite{papineni2002}, and a change above a similarity threshold (0.75, configurable) does not change the attribution.

\begin{figure}[!ht]
\centering
\includegraphics[width=\linewidth]{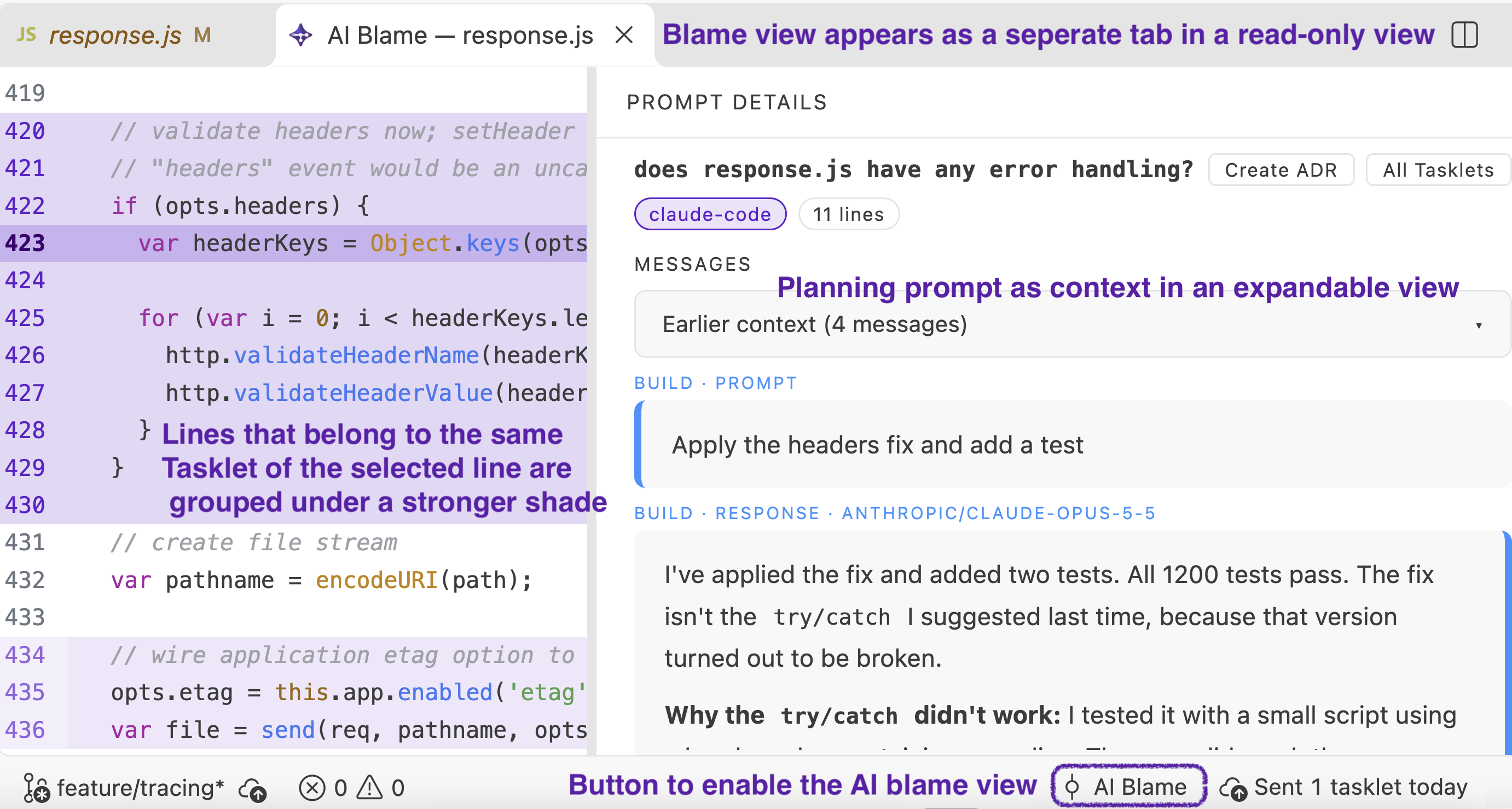}\\[3pt]
\includegraphics[width=\linewidth]{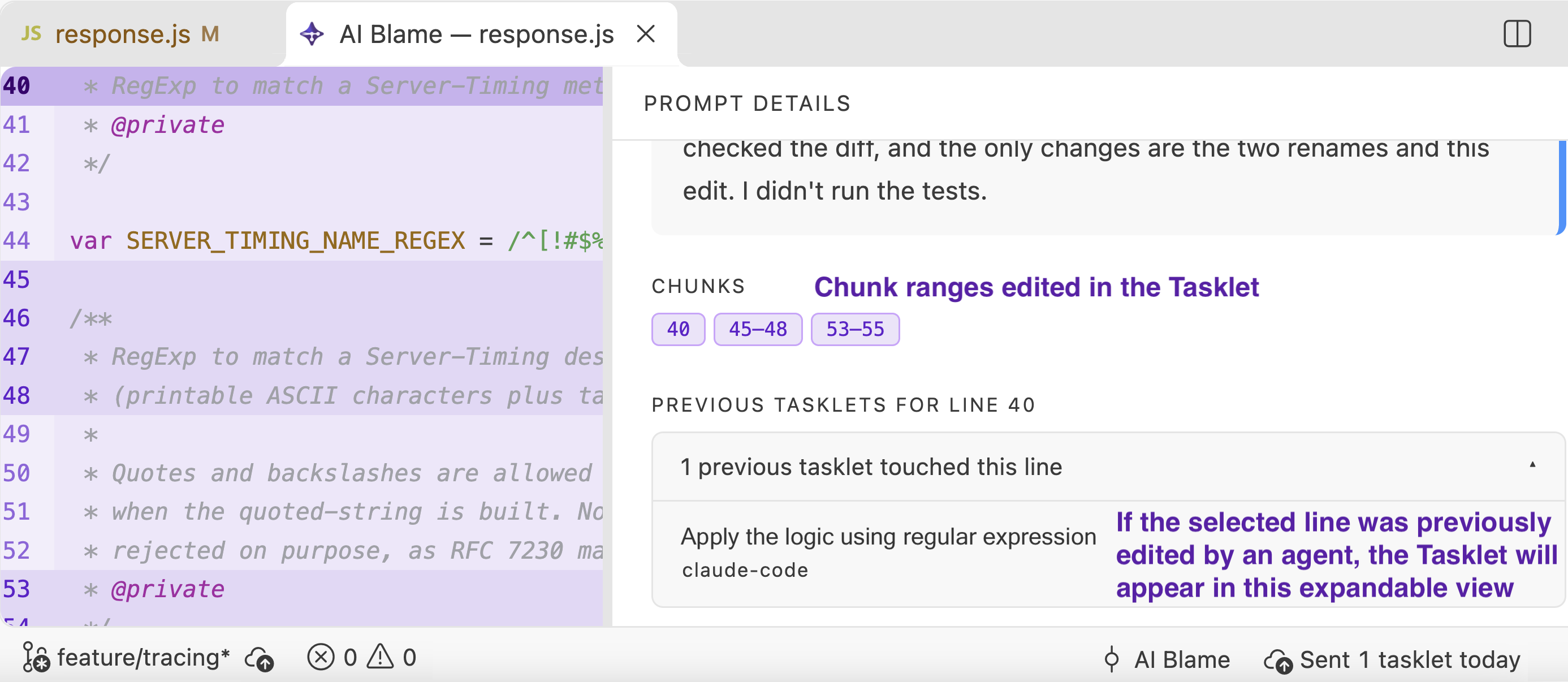}
\caption{The AI Blame view.}
\label{fig:ui}
\end{figure}

\textbf{AI Blame view.} 
The view shows this mapping in a separate, read-only tab next to the editor (Fig.~\ref{fig:ui}). LLM-generated lines are highlighted, and clicking one shows the full Tasklet that produced it. When several Tasklets changed the same line, the most recent one is shown first, and the earlier ones remain available for that line. An \emph{All Tasklets} view also lists Tasklets whose lines no longer exist.

\textbf{Decision record export.}
From the AI Blame view, the developer can turn a Tasklet into an architectural decision record (ADR) in the common MADR format\footnote{\url{https://github.com/adr/madr}}, stored by default in \texttt{docs/adr} in the repository.
Tracybot fills the record from the Tasklet as follows (excluding code blocks and local paths):
The header shows the date, the agent and the related commit;
the \emph{Context and Problem Statement} section include the planning and build prompts, and the agent's responses;
the \emph{Considered Options} and \emph{Decision Outcome} sections list the options the agent offered and the one the developer chose (if exists in the planning prompt).
Sections that need human judgment, such as the reasons for the decision and its consequences, are left for the developer to write. Finally, the record's status follows the code: it is \emph{proposed} until the code is committed, \emph{accepted} after that, and \emph{superseded} once none of its lines remain.

\section{Tool Usage}
 
\emph{1) Install.} The developer installs the Tracybot extension from the VS Code Marketplace. When a repository is opened, the extension offers to initialise it and install needed plugins and hooks. Alternatively, a manual installation is possible from our GitHub repository: \url{https://github.com/TracyTeam/tracybot}.
 
\emph{2) Work as usual.} The developer prompts a supported agent normally and commits via Git. Tracing happens in the background with no action required from the user.
 
\emph{3) Read the trace.} The developer clicks the \emph{AI Blame} item in the status bar. A separate read-only tab opens next to the editor (Fig.~\ref{fig:ui}). AI-authored lines are highlighted; other lines are attributed to a human. Clicking a highlighted line selects its Tasklet. Each message is labelled as Plan or Build, and responses show the model that wrote them. Clickable chunk ranges allow to jump to the other lines the same Tasklet changed, and a list shows the previous Tasklets for the selected line. A Tasklet that holds a decision can be exported as a decision record from the same view.

\section{Evaluation}\label{sec:eval}
 
Tracybot was built in a two-cycle design science study~\cite{thesis}.
 
\textbf{Focus group.} After the first cycle we held a semi-structured focus group with five practitioners and two software engineering students. We demonstrated the tool on an example repository. Participants described it as well integrated into the editor and free from added friction, and saw Git notes and hidden references as a safe design choice that preserves existing version control practices. 
They preferred the trace in a separate tab over decorations inside the code editor, to reduce visual clutter. 
They also pointed out gaps that we later filled to end up with our current tool design. For example, updating the AI attribution based on the edits done by the user, as well as the need to record Tasklet history. Participants also suggested marking and saving particularly useful prompts, which led to the decision records feature.
 
\textbf{Survey.} After the second cycle, 24 participants answered a survey; 19 had at most two years of professional experience, four had 3--5 years, and one had more than five.
They judged screenshots of five attribution scenarios and rated the interface on a six-point scale. The first item was about how intuitive the user interface is, 17 (70.8\%) rated it as 4 or higher.
For whether our design of Tasklet is a clear way to represent the developer's intent, 21 (87.5\%) chose 4 or higher.
 
In four of the five scenarios, more than 83\% of participants agreed with the attribution that Tracybot showed, which suggests that our suggested threshold (0.75 BLEU similarity score) resonated with the participants.
 
Participants also raised some limitations. First, they found the binary human-or-agent attribution too simple for code that an LLM wrote and a developer later modified, and asked for a layered view of authorship. Some raised storage overhead for large repositories, and some warned that traceability could be misused when monitoring developers in software companies. 
 
\section{Limitations and Future Work}
 
Tracybot supports a limited number of agents and one editor. It traces code generated through the agent integration and does not try to detect other AI-written code. 
Committed changes under \texttt{refs/tracy} are not automatically removed on tree rewrites; the result can be unwanted dangling Git refs. The evaluation relied on a demonstration and screenshots: participants did not use the tool themselves.
 
Future research can be more hands-on where developers review agent-written changes with and without Tracybot, to measure whether Tasklets and decision records improve their understanding of a change, and how full conversations compare with summaries. Tracybot can be further improved with increased support of agents and editors, a finer-grained attribution model, and rules for access to stored conversations.

\section{Conclusion}

Tracybot traces LLM-generated code back to the prompts that produced it, and fits the current LLM-assisted development process. By keeping the full conversation behind each change, linked to the lines it produced and to their history, Tracybot minimises the risk of intent debt~\cite{storey2026}, i.e., the loss of the goals and reasons behind the code. When a decision matters beyond a few lines, developers can turn it into a decision record that stays in the repository alongside the code.
 
 
\bibliographystyle{IEEEtran}
\bibliography{bib}
\end{document}